\documentclass[sigconf]{acmart}

\usepackage{balance}

\def\BibTeX{{\rm B\kern-.05em{\sc i\kern-.025em b}\kern-.08emT\kern-.1667em\lower.7ex\hbox{E}\kern-.125emX}}
    
\copyrightyear{2019} 
\acmYear{2019} 
\setcopyright{acmlicensed}
\acmConference[KDD '19]{The 25th ACM SIGKDD Conference on Knowledge Discovery and Data Mining}{August 4--8, 2019}{Anchorage, AK, USA}
\acmBooktitle{The 25th ACM SIGKDD Conference on Knowledge Discovery and Data Mining (KDD '19), August 4--8, 2019, Anchorage, AK, USA}
\acmPrice{15.00}
\acmDOI{10.1145/3292500.3330885}
\acmISBN{978-1-4503-6201-6/19/08}

\usepackage{amsmath,amsthm,amssymb,amsfonts}
\usepackage{graphicx}
\usepackage{xspace}
\usepackage{algorithm}
\usepackage{algpseudocode}
\usepackage{color}
\usepackage{pgfplotstable}
\pgfplotsset{compat=1.5}
\usepackage{tikz}
\usepackage{tabularx}

\usetikzlibrary{shapes, shadows,positioning}
\usepgfplotslibrary{fillbetween}

\algdef{SE}[SUBALG]{Indent}{EndIndent}{}{\algorithmicend\ }%
\algtext*{Indent}
\algtext*{EndIndent}

\algrenewcommand\algorithmiccomment[2][\normalsize]{{#1\hfill\(\triangleright\) \em \textcolor{gray}{#2}}}

\DeclareMathOperator*{\argmax}{arg\,max}

\newcommand{\paragraphbe}[1]{\vspace{0.75ex}\noindent{\bf \em #1}\hspace*{.3em}}

\newcommand{\D}{\mathcal{D}}
\newcommand{\Dref}{\mathcal{D}_\text{ref}}
\newcommand{\Duser}{\mathcal{D}_u}
\newcommand{\Dtrain}{\mathcal{D}_\text{train}}

\newcommand{\Prob}{\mathbf{Pr}}

\renewcommand{\U}{\mathcal{U}}
\newcommand{\Utrain}{\mathcal{U}_\text{train}}
\newcommand{\Uref}{\mathcal{U}_\text{ref}}
\newcommand{\TrainAlgo}{\mathcal{T}_\text{target}}

\newcommand{\FeatExtract}{\texttt{HistogramFeature}}
\newcommand{\Sample}{\texttt{SampleQueries}}

\newcommand{\Dialogue}{Dialogs\xspace}

\newcommand{\faudit}{f_\text{audit}}
\newcommand{\fshadow}{f^\prime}

\makeatletter
\newcommand\fs@betterruled{%
  \def\@fs@cfont{\bfseries}\let\@fs@capt\floatc@ruled
  \def\@fs@pre{\vspace*{5pt}\hrule height.8pt depth0pt \kern2pt}%
  \def\@fs@post{\kern2pt\hrule\relax}%
  \def\@fs@mid{\kern2pt\hrule\kern2pt}%
  \let\@fs@iftopcapt\iftrue}
\floatstyle{betterruled}
\restylefloat{algorithm}
\makeatother

\begin{document}
\title{
Auditing Data Provenance in Text-Generation Models
}

\author{Congzheng Song}
\affiliation{%
  \institution{Cornell University}
}
\email{cs2296@cornell.edu}

\author{Vitaly Shmatikov}
\affiliation{%
  \institution{Cornell Tech}
  }
\email{shmat@cs.cornell.edu}

\begin{abstract}
To help enforce data-protection regulations such as GDPR and detect
unauthorized uses of personal data, we develop a new \emph{model auditing}
technique that helps users check if their data was used to train a machine
learning model.  We focus on auditing deep-learning models that generate
natural-language text, including word prediction and dialog generation.
These models are at the core of popular online services and are often
trained on personal data such as users' messages, searches, chats,
and comments.

We design and evaluate a black-box auditing method that can detect,
with very few queries to a model, if a particular user's texts were used
to train it (among thousands of other users).  We empirically show that
our method can successfully audit well-generalized models that are not
overfitted to the training data.  We also analyze how text-generation
models memorize word sequences and explain why this memorization makes
them amenable to auditing.


\end{abstract}


\maketitle

\section{Introduction}
\label{sec:intro}

Data-protection policies and regulations such as the European Union's
General Data Protection Regulation (GDPR)~\cite{gdpr} give users the right
to know how their data is processed.  As machine learning (ML) becomes
a core component of data processing in many offline and online services,
and incidents such as DeepMind's unauthorized use of NHS patients' data to
train ML models~\cite{deepmind} illustrate the resulting privacy risks,
it is essential to be able to \emph{audit} the provenance of personal
data used for model training.

In this paper, we design and evaluate a technology that can \textbf{help
users audit ML models to determine if their data was used to train
these models}.  We focus specifically on auditing models that generate
natural-language text.  Text-generation models for tasks such as
next-word prediction (the basis of query autocompletion and predictive
virtual keyboards) and dialog generation (the basis of chatbots and
automated customer service) are extensively trained on personal data,
including users' messages, documents, chats, comments, and search
queries.  Our technology can help users audit a publicly available
text-generation model and see if their words were used, perhaps without
their permission, to create this model.  Furthermore, our work sheds new
light on \textbf{how deep learning-based, text-generation models memorize
their training data}\textemdash a topic that has important implications
for both data privacy and natural language processing.

The problem of auditing is closely related to the problem of membership
inference (see Section~\ref{sec:related}), but auditing text-generation
models requires new technical machinery vs.\ membership inference in
image-classification and categorical models.

First, we assume a very restrictive auditing scenario, which we believe
matches how an individual user may audit a deployed ML-based service
in practice.  The auditor has only black-box access to the model and
can query it only on a limited number of inputs.  We assume that the
model's output does not include numeric probabilities or confidence values
(deployed models rarely release these values).  Furthermore, we consider
scenarios where the model's output is restricted to a relatively small
list of words or even a single word.  This precludes the application of
most previously proposed membership inference methods.


Second, we work with text-generation models that are trained on the
data of hundreds or thousands of users and are well-generalized, i.e.,
their accuracy on test inputs is not substantially different from their
accuracy on training inputs.  This precludes the application of membership
inference methods that exploit the test-train accuracy gap exhibited by
overfitted models.

Third, state-of-the-art text-generation models are based on recurrent
neural networks (RNNs).  We investigate how these models overfit to
their training data, what signal this overfitting creates in their
outputs, and how to exploit this signal for effective auditing.
We show that overfitting in text-generation models appears to manifest
primarily via shifted probability distributions over the models'
output space.  Specifically, we show that these models tend to assign
significantly higher rank to relatively rare words when they appear
in a familiar context (e.g., in a sentence seen during training).
This does not affect the top-ranked, likeliest word generated by
the model and therefore\textemdash in contrast to ``conventional''
overfitting\textemdash does not manifest in reduced test accuracy.

Fourth, we show how to use auxiliary public datasets and cross-domain
training when the auditor does not know the distribution from which the
training data for the target model was drawn.

Fifth, we focus on user-level auditing (vs.\ inferring membership of
individual inputs in the training dataset) and measure how many queries
are needed to determine if the user's data was used\textemdash possibly
in combination with the data from thousands of other users\textemdash
to train the model.  We quantitatively show that sequences that include
relatively rare words are more effective for auditing than word sequences
randomly selected from the user's data.  We also measure the robustness
of our auditing methodology to noise and errors in the test inputs used
for auditing.  This is important because the user may not know exactly
which of his chats or online comments were used, or when the model
creator may have started training on the user's data.


Our black-box auditing methodology is very effective.  In our experiments
on the Reddit, SATED, and \Dialogue tasks for, respectively, word
prediction, translation, and dialog generation, it performs perfectly
(i.e., its AUC score is 1) when the models are trained on the data of
hundreds of users and the models' outputs cover the entire vocabulary.
Furthermore, it requires surprisingly few queries.  If the auditor selects
query sequences that include relatively rare words, a \emph{single query}
achieves AUC between 0.8 and 0.9 depending on the task, and 8 queries
achieve almost perfect AUC.

If the word-prediction and dialog-generation models are restricted to
generate and rank only the 500 likeliest words, AUC score of our auditor
remains above 0.9.  If the translation model generates a \emph{single
word} (as opposed to a ranked list of words), the auditor can still infer
with a much-better-than-random probability if the model was trained on
the word sequences of a particular user.  For the Reddit word-prediction
model, the auditor's AUC score remains close to 0.9 even if the model was
trained on the data of over 4,000 users.  Furthermore, we empirically
show that our auditing is robust to a significant amount of noise and
errors in the audit queries.  These results demonstrate that auditing
modern text-generation models is feasible in realistic scenarios.

Finally, to explain why auditing works, we provide new insights into
memorization in different types of text-generation models.  For example,
we demonstrate that deep learning-based translation models are more prone
than the word-prediction models to memorize training sequences in their
inner units.


\section{Background}

\subsection{Deep learning}


A deep learning model is a function $f_\theta$ :
$\mathcal{X}\mapsto\mathcal{Y}$ parameterized by $\theta$, where
$\mathcal{X}$ is the input space and $\mathcal{Y}$ is the output
space.  Supervised training of a model $f_\theta$ aims to find the
best set of parameters $\theta$ using a labeled training dataset $D =
\{(x_i,y_i)\}_{i=1}^n$ and a loss function $L$.


For ML tasks where the input space is discrete and sparse (e.g.,
text or location data), the standard approach is to transform discrete
inputs into a lower-dimensional continuous vector representation.  For a
text corpus with vocabulary $V$, an \emph{embedding} is a function $E:
V\mapsto \mathbb{R}^{d_\text{emb}}$ where $d_\text{emb}$, the dimension
of the embedding, is a hyper-parameter.  In many NLP tasks, the input
is a variable-length sequence of tokens $x=[x^1,\dots, x^l]$ in the
embedding space.  The output $y$ can be either a class label (e.g.,
for sentiment analysis), or a token (e.g., for next-word prediction),
or a sequence of tokens (e.g., for machine translation).




\paragraphbe{Recurrent neural networks} (RNNs) are a common architecture
for text-generation tasks such as next-word prediction.  An RNN maps the
input sequence to a sequence of hidden representations $a=[a^1,\dots,
a^l]$, where the computation of $a^j$ is recursively dependent on the
previous hidden representation $a^{j-1}$ and the current input token
$x^j$, and feeds these hidden representations to a classifier.



\paragraphbe{Sequence-to-sequence models} are a common architecture for
text-generation tasks where both the input $x=[x^1, \dots, x^l]$ and
the output  $y=[y^1, \dots, y^t]$ are sequences of tokens.  A typical
sequence-to-sequence model consists of an encoder RNN and a decoder RNN.
The encoder learns the representation for the input texts, then passes
this representation as the initial state for the decoder, which makes
word predictions one at a time.  Translation models are similar: the
decoder predicts words in the target language by feeding its hidden
representations to a classifier.


\subsection{Text-generation models}
\label{sec:back-language}


\noindent
\textbf{Next-word prediction} is used in many natural-language
applications, including predictive virtual keyboards and query
autocompletion.  Given an input sequence $x= [x^1,\dots, x^{l}]$, the
task is to predict the next token $x^j$ from the context $[x^1,\dots,
x^{j-1}]$.  RNNs are commonly used for this task.  RNN feeds the last
hidden representation $a^{j-1}$ in the context sequence to a $|V|$-way
classifier to predict the next token, where $V$ is the vocabulary.

\paragraphbe{Neural machine translation} (NMT) models based on RNNs
reach near-human performance on many language pairs~\cite{wu2016google}.
The input to these models is a sequence of tokens from the source
language, the output is a sequence of tokens from the target language.
NMT models use the sequence-to-sequence framework.  The input text
is encoded as a hidden representation, and the decoder RNN predicts
translated tokens based on this representation.


\paragraphbe{Dialog generation} aims to generate replies in a
conversation.  It is a common component of chatbots and question-answering
services.  The input is a sentence, the output is the next sentence
in the same conversation.  Dialog-generation models can also employ a
sequence-to-sequence architecture~\cite{vinyals2015neural, li2016persona}.
Similar to NMT, the model encodes the input sentence to a hidden
representation, then generates the reply by passing this representation
to the decoder.

\paragraphbe{Loss functions.} 
For the next-word prediction task, given an input sequence $x=[x^1, \dots,
x^l]$, the RNN models the conditional probability
$\Prob(x^j|x^1,\dots,x^{j-1}) = f(x^1,\dots,x^{j-1})$ 
and aims to maximize the probability for the sequence 
$\Prob(x) = \prod_{j=1}^l \Prob(x^j|x^1,\dots,x^{j-1})$.  
The loss function used when training the model is thus
the negative log likelihood:
$L(f(x), x) = -\sum_{j=1}^l \log f(x^1,\dots,x^{j-1})$.
For the machine translation and dialog-generation tasks where the input is
$x$ and the target is $y=[y^1, \dots, y^t]$, the sequence-to-sequence
model computes the probability $\Prob(y^j|y^1,\dots,y^{j-1};x)$ as
$f(y^1,\dots,y^{j-1};x)$.  Similar to the next-word prediction task,
the loss function is the negative log probability on the target sequence.

\begin{figure*}[!t]
\centering
\includegraphics[width=0.9\textwidth]{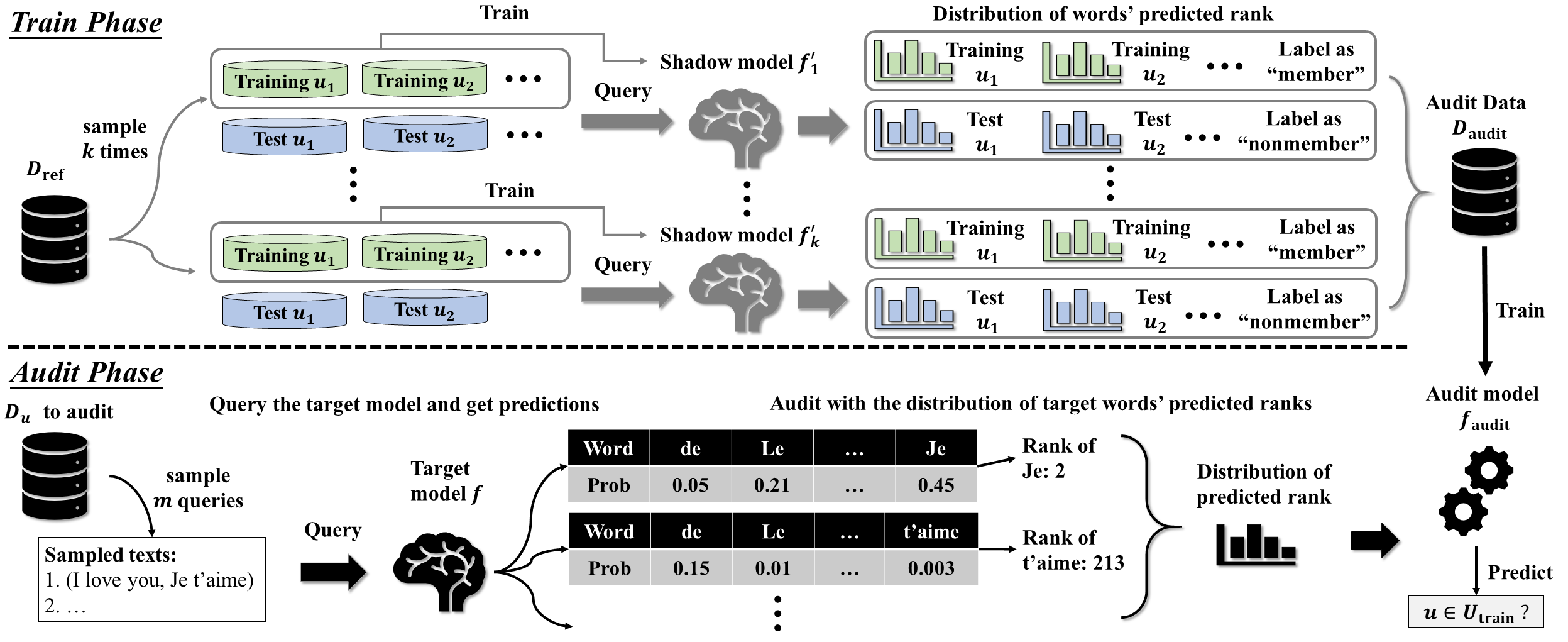}
\caption{Overview of the auditing process.  In the \emph{Train} phase,
the auditor trains an audit model; in the \emph{Audit} phase, he applies
the audit model to infer if the user's data is part of the target's
training dataset.}
\vspace{-2ex}
\label{fig:overview}
\end{figure*}


\section{Auditing text-generation models}

\label{problem-stmt}


Consider a training dataset $\Dtrain$ where each row is associated
with an individual user, and let $\Utrain$ be the set of all users
in $\Dtrain$.  The target model $f$ is trained on $\Dtrain$ using a
training protocol $\TrainAlgo$, which includes the learning algorithm
and the hyper-parameters that govern the training regime.  As described
in Section~\ref{sec:back-language}, a text-generation model $f$ takes
as input a sequence of tokens $x$ and outputs a prediction $f(x)$
for a single token (if the task is next-word prediction) or a sequence
of tokens (if the task is machine translation or dialog generation).
The prediction $f(x)$ is a probability distribution or a sequence of
distributions over the training vocabulary $V$ or a subset of $V$.
We assume that the tokens in the model's output space are ranked (i.e.,
the output distribution imposes an order on all possible tokens) but
do not assume that the numeric probabilities from which the ranks are
computed are available as part of the model's output.


The goal of \textbf{auditing} is to infer user-level membership against
the target model $f$, i.e., to decide whether a user $u\in \Utrain$
or not.

We assume that the auditor has black-box access to $f$: given an input
query $x$, the auditor can observe $f(x)$.  In realistic deployments of
text-generation models, the auditor may not be able to observe the entire
vector of ranked words $f(x)$ but only several top-ranked predictions.
In our experiments in Section~\ref{sec:outputsize}, we vary the size
of the model's output and show how it affects the accuracy of auditing.

We assume that the auditor knows the learning algorithm used to
create $f$ but he may or may not know the training hyper-parameters
(see Section~\ref{sec:diffhyper}).  The auditor also needs an auxiliary
dataset $\Dref$ to train shadow models that perform the same task as $f$.



\label{sec:shadow}

Fig~\ref{fig:overview} outlines the auditing process.  Similar to standard
membership inference~\cite{shokri2017membership}, the auditor's goal
is to learn to distinguish the outputs produced by the target model
on sequences that it trained on and its outputs on sequences that it
did not see during training.  For this purpose, the auditor builds a
binary user-level membership classifier $\faudit$ that takes as input a
(processed) list of predictions obtained by querying $f$ with a subset
of the user's dataset $\Duser$ and outputs a decision on $u\in \Utrain$.
In Section~\ref{sec:selection}, we show that a small subset of $\Duser$
is sufficient for this purpose.


\paragraphbe{Training shadow models.}
To collect the data for training $\faudit$, the auditor first trains $k$
shadow models $\fshadow_1\dots\fshadow_k$ (that ``simulate'' $f$) using
the same protocol $\TrainAlgo$ as $f$ with the same hyper-parameters (if
known) or varying the hyper-parameters as in Section~\ref{sec:diffhyper}.


The training data for each shadow is a random user subset
$\Uref^\text{train}\in\Uref$ of the auxiliary dataset $\Dref$.  Our shadow
training technique is inspired by~\cite{shokri2017membership}, but one
essential distinction is that in our case the shadow-training data does
not need to be drawn from the same distribution as the training data of
the target model In Section~\ref{crossdomain}, we show that public sources
can be used for $\Dref$ and the loss in audit accuracy is negligible
when $\Dtrain$ and $\Dref$ are drawn from different domains.  This is
important for real-world auditing because in practice the auditor may
not know the entire distribution of the target model's training data,
and API limits may prevent the auditor from querying the target model
repeatedly to extract sufficient data for training shadow models as
in~\cite{shokri2017membership}.

The auditor then queries the shadow models with $\D_{\text{ref}, u}$
for each $u$ in $\Uref$ and labels the resulting outputs as ``member''
if $u$ was part of the shadow's training data, ``non-member'' otherwise.
The next step is to use these labeled predictions to train a binary
membership classifier.



\paragraphbe{Training the audit model.}
\label{sec:audittrain}
Record-level membership inference typically uses the output probability
distribution directly as the feature to distinguish between members and
non-members.  User-level membership inference in text-generation models
calls for a different approach.  Each user is associated with multiple
sequences, each of which has multiple words.  Therefore, the auditor can
obtain a collection of output predictions.  On the negative side, the
actual probabilities associated with each prediction may not be available.

As mentioned before, the output prediction $f(x)$ for a input $x$ is a
probability distribution across the entire training vocabulary $V$, i.e.,
a $|V|$-dimensional probability vector.  $|V|$ is generally large and
the probability values are noisy.  Instead of the raw probability values,
we use the \emph{ranks} of the target words in the output distributions
as signals for inferring user-level membership.  As we will show in
Section~\ref{sec:overfit}, even for a well-generalized model (i.e.,
whose test-train accuracy gap is small), there is a substantial gap in
the predicted rank of the same word when it appears in a training text
and a test text.  Specifically, the model ranks relatively rare words
much higher when it sees them during testing in the same context as it
saw them during training.

Given a user $u$'s data $\D_{\text{ref}, u}$, the auditor queries
the shadow model on each data point $(x, y)\in\D_{\text{ref}, u}$
and collects the ranks of $y$ in $f(x)$ into a rank set $R_u$.  Taking
English-to-French machine translation task as an example where $(x, y)$ =
(I love you, Je t'aime), $f(x)=[f(x)^1, f(x)^2]$ is a sequence of two
probability vectors for tokens ``Je'' and ``t'aime.''  The auditor
collects the rank of the probability of ``Je'' in $f(x)^1$ (e.g.,
2) and the rank of the probability of ``t'aime'' in $f(x)^2$ (e.g.,
213), and adds \{2, 213\} to the rank set $R_u$.  Rank 2 means that
the word is the second likeliest prediction in the entire vocabulary.
After collecting the ranks for all $(x, y)\in\D_{\text{ref}, u}$, the
auditor builds a histogram for $R_u$ with a fixed number of bins $d$.
The final feature vector $h_u$ is a $d$-way count vector where each
entry is the count of the ranks in that bin.

The auditor extracts features $h_u$ and labels them as 1 if
$u\in\Uref^\text{train}$ and 0 otherwise.  The auditor repeats this
procedure for each user in each shadow model and obtains a collection of
labeled feature vectors $\D_\text{audit}$.  Finally, the auditor trains
a binary membership classifier $f_\text{audit}$ on $\D_\text{audit}$.
We refer to $f_\text{audit}$ as the \textbf{audit model}.



\paragraphbe{Auditing membership in the training data.}
\label{sec:inference}
At inference (i.e., audit) time, the auditor queries the target model
$f$ with the user's data $\Duser$.  If the number of queries to $f$ is
limited, only a sample from $\Duser$ is used.  It can be random, but we
show in Section~\ref{sec:selection} that it is more effective to select
test inputs that have the smallest frequency counts in their labels $y$,
i.e., sequences with relatively rare words are more useful for auditing.


After querying $f$, the auditor processes the corresponding outputs and
obtains a feature vector $h_u$ that describes the distribution of the
predicted ranks for each word in $\Duser$.  Finally, the auditor feeds
$h_u$ to $\faudit$, which decides whether $u\in \Utrain$ or not.

\section{Experiments}
\label{sec:experiments}

\subsection{Datasets}

The \textbf{Reddit comments dataset} (Reddit) is a
randomly chosen month (November 2017) from the public Reddit
dataset.\footnote{\url{1https://bigquery.cloud.google.com/dataset/fh-bigquery:reddit
comments }}  We filtered it to retain only the users with at least 150 but
no more than 500 posts, for a total of 83,293 users with 247 posts each on
average.  We use the resulting dataset for the next-word prediction task.

The \textbf{speaker annotated TED talks dataset} (SATED) consists of
transcripts from TED talks,\footnote{\url{https://www.ted.com/talks}}
totaling 2,324 talks with roughly 271K sentences in each
language~\cite{michel2018extreme}.  The dataset contains English-French
(en-fr), English-German (en-de) and English-Spanish (en-es) language
pairs and speaker annotation.  We use the data from the en-fr pair for
the machine translation task.

The \textbf{Cornell movie dialogs corpus} (\Dialogue) is
a collection of fictional conversations extracted from movie
scripts~\cite{Danescu-Niculescu-Mizil+Lee:11a}.  There are a total of
220,579 exchanges between pairs of characters engaging in at least 5
exchanges, involving 9,035 characters from 617 movies.  We use this
dataset for the dialogue-generation task.

\paragraphbe{Cross-domain reference datasets.}
The auditor may not know the distribution on which the
target model was trained and thus needs a reference dataset
to train its shadow models.  In our experiments, we use
public datasets for this purpose.  As the cross-domain
reference dataset for word prediction, we use the Wikitext-103
corpus\footnote{\url{https://einstein.ai/research/blog/the\%2Dwikitext\%2Dlong\%2Dterm\%2Ddependency\%2Dlanguage\%2Dmodeling\%2Ddataset}}
obtained by a Wikipedia crawl.  For translation, we use the
English-French pair in the Europarl dataset~\cite{koehn2005europarl},
a parallel language corpus extracted from the proceedings of the
European Parliament.  For dialog generation, we use the Ubuntu dialogs
dataset~\cite{lowe2015ubuntu}, which contains two-person technical
support chat logs.

These datasets are not labeled with individual users, thus we split them
into random $n_u$ subsets, each corresponding to an artificial ``user.''
Our experiments show that we can produce effective audit models even
with this artificial separation into users and even though the topics
of the reference datasets are very different from the target models'
training datasets (e.g., technical support chats vs.\ conversations
between movie characters).

\subsection{Performance of target models}
\label{sec:target_performance}

\begin{table}[t!]
\footnotesize
\centering
\caption{Performance of target models.  Acc is word prediction accuracy,
perp is perplexity.}
\begin{tabular}{l|l|rr|rr}
Dataset & Model & Train Acc & Test Acc & Train Perp & Test Perp\\ \hline\hline
Reddit & 1-layer LSTM~\cite{hochreiter1997long} & 0.184 & 0.206 & 102.22 & 113.14 \\ 
SATED & Seq2Seq w/ attn~\cite{michel2018extreme} & 0.587 &	0.535 &	6.36 & 10.28 \\
\Dialogue & Seq2Seq w/o attn & 0.283 & 0.264 & 45.57 & 61.11 \\
\hline 
\end{tabular}
\label{tbl:target_performance}
\end{table}


We use standard architectures and hyper-parameters to train target models
(see Section~\ref{sec:setup}) and evaluate their performance using
\emph{word prediction accuracy} ~$=\frac{1}{M}\sum_{i=1}^n\sum_{j=1}^{l_i} \mathbb{I}(\argmax f(x_i)^j = y_i^j) $ and \emph{perplexity}~ $=2^{-\frac{1}{M}\sum_{i=1}^n\sum_{j=1}^{l_i}\log f(x_i)^j[y_i^j]}$,
 where $n$ is the number of data points, $M=\sum_i l_i$ the sum of the number of tokens in all labels,
$\mathbb{I}$ is the indicator function that outputs 1 if the predicted
token $\argmax f(x_i)^j$ equals the label token $y_i^j$ and 0 otherwise,
and $f(x_i)^j[y_i^j]$ is the probability of predicting $y_i^j$ in
$f(x_i)^j$.  Perplexity is measured as 2 to the power of the entropy of
the label predictions.  The lower the perplexity, the better the model
fits the data.

Table~\ref{tbl:target_performance} shows the results for models
trained on 300 users, with the test data sampled from 300 disjoint
users from the training set.  These results match the literature.
On Reddit, test accuracy of word prediction is 20\%, similar
to~\cite{mcmahan2017learning}.  On SATED, test perplexity is 10, close
to~\cite{luong2015deep}.  Low test perplexity shows that the models are
learning a meaningful language-generation process.  Test-train accuracy
gaps are below 5\%, indicating that the models are not overfitted.
Perplexity gaps are within 15, which is relatively small.

\subsection{Performance of auditing}

To train shadow models, we sample a set of ``shadow users'' disjoint from
both the training and test users.  The number of shadow users is twice the
number of training users.  We use one half of the shadow users to train
shadow models and the other half to collect the shadow models' outputs on
the non-members of their training datasets (see Section~\ref{sec:shadow}).
We train 10 shadow models for all tasks and use a linear SVM as the
audit classifier.

Our metrics are precision (the percentage of users classified by
the audit model as ``members'' who are indeed members), recall (the
percentage of members who are classified as ``members''), accuracy
(the percentage of all users who are classified correctly), and AUC,
the area under the ROC curve that shows the gap between the scores (i.e.,
distances to the decision hyperplane of SVM) given by the audit model to
members and non-members.  We use 300 members and non-members.  Therefore,
the baseline for all metrics is 0.5, corresponding to random guessing.


\textbf{Our audit model achieves the perfect score} (i.e., 1) on all
metrics for all datasets and models when there is no restriction on the
output size of the target models (i.e., they produce predictions over
the entire vocabulary) and the auditor can query the target models any
number of times.

\paragraphbe{Effect of different hyper-parameters.}
\label{sec:diffhyper}
To demonstrate that knowledge of the target model's hyper-parameters
is not essential for successful auditing, we train 10 shadow models
for each task with different training configurations (detailed in
Appendix~\ref{sec:config}).  Table~\ref{tbl:diffhyper} shows the results.
Auditing scores are still above 0.95 on nearly all metrics for all tasks
and models.


\begin{table}
\caption{Effect of training shadow models with different hyper-parameters
than the target model.}
\label{tbl:diffhyper}
\centering
\footnotesize
\begin{tabular}{l|rrrr}
     Dataset & Accuracy & AUC & Precision & Recall \\ \hline\hline
     Reddit & 0.990 & 0.993 & 0.983 & 0.996 \\
     SATED & 0.965 & 0.981 & 0.937 & 0.996 \\
     Dialogs & 0.978 & 0.998 & 0.958 & 1.000 \\ \hline 
\end{tabular}
\end{table}

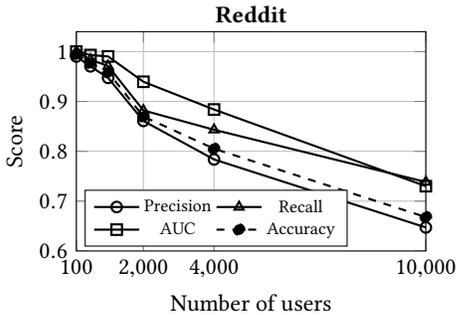
\begin{figure}[t!]
\begin{tikzpicture}
\begin{axis}[title style={yshift=-1.5ex}, title=\textbf{Reddit}, xlabel=Number of users,ylabel=Score, 
ymin=0.6, xmin=100, xmax=10000, xtick={100, 2000, 4000, 10000}, 
scaled x ticks=false,
legend style = {nodes={scale=0.75}, at = {(0.4,0.02)}, legend columns=2, anchor=south},
width=0.35\textwidth, height=4.5cm, grid=both]
\pgfplotstableread{plot/num_users.txt}\mydata;
\addplot[thick, mark=o] table
[x expr=\thisrow{p},y expr=\thisrow{pre}] {\mydata};
\addlegendentry{Precision};
\addplot[thick, mark=triangle] table
[x expr=\thisrow{p},y expr=\thisrow{rec}] {\mydata};
\addlegendentry{Recall};
\addplot[thick, mark=square] table
[x expr=\thisrow{p},y expr=\thisrow{auc}] {\mydata};
\addlegendentry{AUC};
\addplot[thick, dashed,mark=*] table
[x expr=\thisrow{p},y expr=\thisrow{acc}] {\mydata};
\addlegendentry{Accuracy};
\end{axis}
\end{tikzpicture} 
\caption{Effect of the number of Reddit users used to train a
word-prediction model.}
\label{fig:numusers}
\end{figure}

\paragraphbe{Effect of the number of users.}
To evaluate how the number of users in the training dataset affects
the auditor's ability to infer the presence of a single user, we train
word-prediction models on 100, 500, 1,000, 2000, 4,000, and 10,000
users from the Reddit dataset.  Test users and shadow users are disjoint
samples of the same size.

Fig.~\ref{fig:numusers} shows the results.  When the number of users is
under 1,000, all metrics are at least 0.95.  With 4,000 users, precision
drops below 0.8 while AUC is still around 0.9.  Audit performance drops
more significantly when the number of users is 10,000.


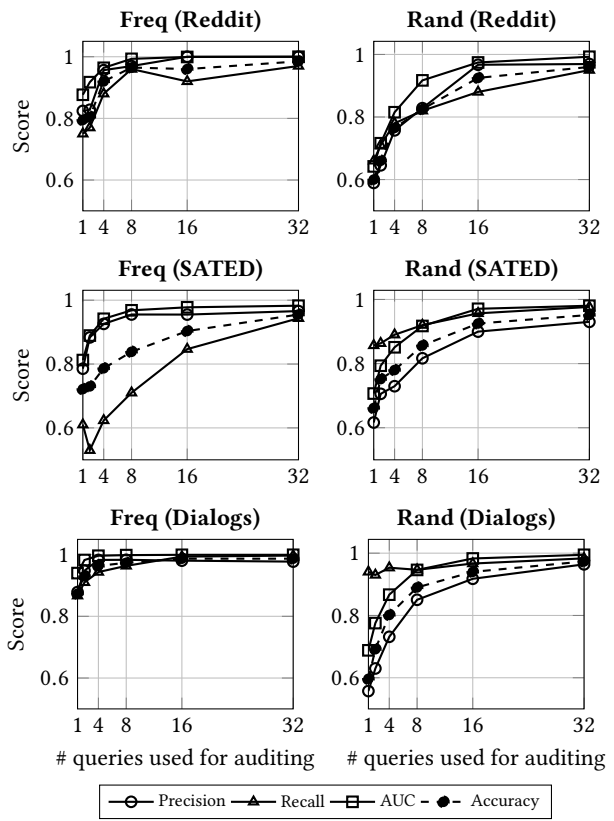
\begin{figure}[t!]
\centering
\begin{tabular}{cc}
\begin{tikzpicture}
\begin{axis}[title style={yshift=-1.5ex}, title=\textbf{Freq (Reddit)}, ylabel=Score, ymin=0.5, xmin=1, xmax=32,  xtick={1,4,8,16,32}, width=0.25\textwidth, grid=both, name=ax1]
\pgfplotstableread{plot/reddit_rare_rand.txt}\mydata;
\addplot[thick, mark=o] table
[x expr=\thisrow{p},y expr=\thisrow{pre1}] {\mydata};
\addplot[thick, mark=triangle] table
[x expr=\thisrow{p},y expr=\thisrow{rec1}] {\mydata};
\addplot[thick, mark=square] table
[x expr=\thisrow{p},y expr=\thisrow{auc1}] {\mydata};
\addplot[thick, dashed,mark=*] table
[x expr=\thisrow{p},y expr=\thisrow{acc1}] {\mydata};
\end{axis}

\begin{axis}[title style={yshift=-1.5ex}, title=\textbf{Rand (Reddit)}, ymin=0.5, xmin=1, xmax=32, xtick={1,4,8,16,32},
width=0.25\textwidth, grid=both, at={(ax1.south east)}, xshift=1cm,]
\pgfplotstableread{plot/reddit_rare_rand.txt}\mydata;
\addplot[thick, mark=o] table
[x expr=\thisrow{p},y expr=\thisrow{pre2}] {\mydata};
\addplot[thick, mark=triangle] table
[x expr=\thisrow{p},y expr=\thisrow{rec2}] {\mydata};
\addplot[thick, mark=square] table
[x expr=\thisrow{p},y expr=\thisrow{auc2}] {\mydata};
\addplot[thick, dashed,mark=*] table
[x expr=\thisrow{p},y expr=\thisrow{acc2}] {\mydata};
\end{axis}

\end{tikzpicture} 

\\
\begin{tikzpicture}

\begin{axis}[title style={yshift=-1.5ex}, title=\textbf{Freq (SATED)},  ymin=0.5, xmin=1, xmax=32, xtick={1,4,8,16,32}, ylabel=Score,
width=0.25\textwidth, grid=both, name=ax1]
\pgfplotstableread{plot/sated_rare_rand.txt}\mydata;
\addplot[thick, mark=o] table
[x expr=\thisrow{p},y expr=\thisrow{pre1}] {\mydata};
\addplot[thick, mark=triangle] table
[x expr=\thisrow{p},y expr=\thisrow{rec1}] {\mydata};
\addplot[thick, mark=square] table
[x expr=\thisrow{p},y expr=\thisrow{auc1}] {\mydata};
\addplot[thick, dashed,mark=*] table
[x expr=\thisrow{p},y expr=\thisrow{acc1}] {\mydata};
\end{axis}

\begin{axis}[title style={yshift=-1.5ex}, title=\textbf{Rand (SATED)}, ymin=0.5, xmin=1, xmax=32, xtick={1,4,8,16,32},
width=0.25\textwidth, grid=both,  at={(ax1.south east)}, xshift=1cm]
\pgfplotstableread{plot/sated_rare_rand.txt}\mydata;
\addplot[thick, mark=o] table
[x expr=\thisrow{p},y expr=\thisrow{pre2}] {\mydata};
\addplot[thick, mark=triangle] table
[x expr=\thisrow{p},y expr=\thisrow{rec2}] {\mydata};
\addplot[thick, mark=square] table
[x expr=\thisrow{p},y expr=\thisrow{auc2}] {\mydata};
\addplot[thick, dashed,mark=*] table
[x expr=\thisrow{p},y expr=\thisrow{acc2}] {\mydata};
\end{axis}
\end{tikzpicture} 
\\
\begin{tikzpicture}
\begin{axis}[title style={yshift=-1.5ex}, title=\textbf{Freq (\Dialogue)}, xlabel=\#  queries used for auditing, ymin=0.5, xmin=1, xmax=32,xtick={1,4,8,16,32},  ylabel=Score,
width=0.25\textwidth, grid=both, name=ax1, 
 legend style = {nodes={scale=0.75}, at = {(1.15,-0.65)}, legend columns=4, anchor=south}]
\pgfplotstableread{plot/cornell_rare_rand.txt}\mydata;
\addplot[thick, mark=o] table
[x expr=\thisrow{p},y expr=\thisrow{pre1}] {\mydata};
\addlegendentry{Precision};
\addplot[thick, mark=triangle] table
[x expr=\thisrow{p},y expr=\thisrow{rec1}] {\mydata};
\addlegendentry{Recall};
\addplot[thick, mark=square] table
[x expr=\thisrow{p},y expr=\thisrow{auc1}] {\mydata};
\addlegendentry{AUC};
\addplot[thick, dashed,mark=*] table
[x expr=\thisrow{p},y expr=\thisrow{acc1}] {\mydata};
\addlegendentry{Accuracy};
\end{axis}

\begin{axis}[title style={yshift=-1.5ex}, title=\textbf{Rand (\Dialogue)}, xlabel=\# queries used for auditing, ymin=0.5, xmin=1, xmax=32, xtick={1,4,8,16,32},
width=0.25\textwidth, grid=both,at={(ax1.south east)}, xshift=1cm,]
\pgfplotstableread{plot/cornell_rare_rand.txt}\mydata;
\addplot[thick, mark=o] table
[x expr=\thisrow{p},y expr=\thisrow{pre2}] {\mydata};
\addplot[thick, mark=triangle] table
[x expr=\thisrow{p},y expr=\thisrow{rec2}] {\mydata};
\addplot[thick, mark=square] table
[x expr=\thisrow{p},y expr=\thisrow{auc2}] {\mydata};
\addplot[thick, dashed,mark=*] table
[x expr=\thisrow{p},y expr=\thisrow{acc2}] {\mydata};
\end{axis}
\end{tikzpicture}
\end{tabular}
\caption{Effect of the number of queries and sampling strategy.  Plots on
the left show the results when the auditor samples the user's data for
queries in the ascending order of frequency counts of tokens in the label;
plots on the right show the results with randomly sampled data.}
\label{fig:numdata}
\end{figure}

\paragraphbe{Effect of the number and selection of audit queries.}
\label{sec:selection}
To measure the performance of auditing when the auditor is restricted
to only a few queries, we vary the number of audit queries between 1,
2, 4, 8, 16, and 32 word sequences.


Fig~\ref{fig:numdata} shows the results.  With 32 queries, audit
performance exceeds 0.9 on all metrics for all datasets.  If query
selection is random, audit performance is low with fewer than 8 queries.
If the auditor queries the target with the user's word sequences whose
summary word-frequency counts are the lowest, \textbf{even with a single
query, the auditor can accurately determine if the user's data was used
to train the model} on the Reddit or \Dialogue dataset.  This remarkable
result demonstrates the extent to which text-generation models memorize
word sequences they were trained on, especially those that contain
relatively rare words.

\begin{table}[t!]
\footnotesize
\centering
\caption{Effect of the model's output size. $|f(x)|$ is the number of words
ranked by $f$.}  
\begin{tabular}{r|rrrr|rrrr}
\textbf{Reddit} & \multicolumn{4}{c|}{\bf Same domain} & \multicolumn{4}{c}{\bf Cross domain}\\
$|f(x)|$ & Acc & AUC & Pre & Rec & Acc & AUC & Pre & Rec \\ \hline\hline
1 & 0.545 & 0.549 & 0.574 & 0.350 & 0.505 & 0.589 & 0.667 & 0.020 \\ 
5 & 0.550 & 0.572 & 0.553 & 0.520 & 0.490 & 0.525 & 0.495 & 0.920 \\ 
10 & 0.580 & 0.602 & 0.582 & 0.570 & 0.500 & 0.552 & 0.500 & 0.950 \\ 
50 & 0.605 & 0.648 & 0.606 & 0.600 & 0.505 & 0.659 & 0.503 & 0.980 \\ 
100 & 0.725 & 0.788 & 0.765 & 0.650 & 0.585 & 0.714 & 0.549 & 0.950 \\ 
500 & 0.970 & 0.998 & 0.970 & 0.970 & 0.905 & 0.992 & 0.988 & 0.820 \\ 
1000 & 0.985 & 0.999 & 0.971 & 1.000 & 0.910 & 0.999 & 1.000 & 0.820 \\ \hline 
\multicolumn{9}{c}{} \\
\textbf{SATED} & \multicolumn{4}{c}{} & \multicolumn{4}{c}{}\\
$|f(x)|$ & Acc & AUC & Pre & Rec & Acc & AUC & Pre & Rec \\ \hline\hline
1 & 0.723 & 0.785 & 0.770 & 0.637 & 0.723 & 0.785 & 0.712 & 0.750 \\ 
5 & 0.748 & 0.838 & 0.767 & 0.713 & 0.767 & 0.834 & 0.755 & 0.790 \\ 
10 & 0.800 & 0.880 & 0.783 & 0.830 & 0.805 & 0.878 & 0.814 & 0.790 \\ 
50 & 0.928 & 0.973 & 0.908 & 0.953 & 0.925 & 0.979 & 0.947 & 0.900 \\ 
100 & 0.948 & 0.981 & 0.944 & 0.953 & 0.942 & 0.978 & 0.965 & 0.917 \\ 
500 & 0.972 & 0.988 & 0.958 & 0.987 & 0.970 & 0.988 & 0.983 & 0.957 \\ 
1000 & 0.960 & 0.984 & 0.939 & 0.983 & 0.967 & 0.985 & 0.973 & 0.960 \\ \hline 
\multicolumn{9}{c}{} \\
\textbf{Dialogs} & \multicolumn{4}{c}{} & \multicolumn{4}{c}{}\\
$|f(x)|$ & Acc & AUC & Pre & Rec & Acc & AUC & Pre & Rec \\ \hline\hline
1 & 0.577 & 0.618 & 0.582 & 0.547 & 0.538 & 0.618 & 0.520 & 0.977 \\ 
5 & 0.575 & 0.642 & 0.582 & 0.530 & 0.552 & 0.643 & 0.528 & 0.970 \\ 
10 & 0.583 & 0.645 & 0.591 & 0.543 & 0.543 & 0.638 & 0.523 & 0.977 \\ 
50 & 0.605 & 0.660 & 0.611 & 0.580 & 0.537 & 0.610 & 0.520 & 0.963 \\ 
100 & 0.647 & 0.714 & 0.643 & 0.660 & 0.570 & 0.669 & 0.541 & 0.920 \\ 
500 & 0.935 & 0.975 & 0.917 & 0.957 & 0.925 & 0.969 & 0.895 & 0.963 \\ 
1000 & 0.972 & 0.995 & 0.955 & 0.990 & 0.962 & 0.992 & 0.948 & 0.977 \\  \hline
\end{tabular}
\label{tbl:outputspace}
\end{table}

\paragraphbe{Effect of the size of the model's output.}
\label{sec:outputsize}
\label{crossdomain}
In a realistic deployment of a text-generation model, its output may
be limited to a few top-ranked words rather than the entire ranked
vocabulary.  We constrain the model's output to the top-ranked 1, 5,
50, 500, and 1000 words, while the other hyper-parameters remain as
in Section~\ref{sec:setup}.  When building the histogram feature vector
for training the audit model (see Section~\ref{sec:audittrain}), we add
an additional feature that counts how many times the ground-truth words
are not among the top predictions output by the model.

Table~\ref{tbl:outputspace} shows the results.  On Reddit and \Dialogue,
the auditor's performance is close to random guessing when the model's
outputs are limited to the top 50 or fewer words, increasing to above
0.9 when the output size is the top 500 words (only 10\% of the entire
vocabulary)\textemdash regardless of whether the shadow models are
trained on the same domain as the target model or a different domain.

For the translation task, \textbf{audit performance is much higher than
random guessing even if the model outputs just one top-ranked word}
and exceeds 0.9 when the model outputs 50 top-ranked words (1\% of
the vocabulary).  These results demonstrate the remarkable extent to
which translation models memorize specific word sequences encountered
in training.

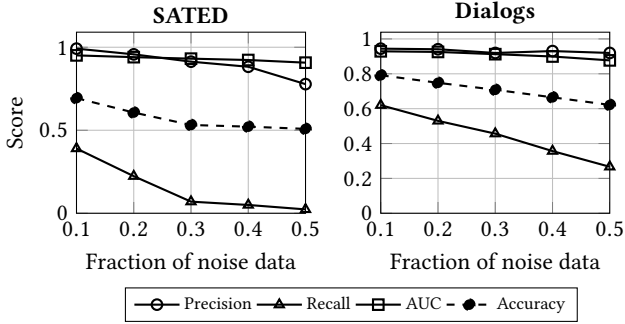
\begin{figure}[t!]
\centering
\begin{tikzpicture}
\begin{axis}[title style={yshift=-1.5ex}, title=\textbf{SATED}, xlabel=Fraction of noise data,ylabel=Score, ymin=0., xmin=0.1, xmax=0.5, xtick=data, legend style = {nodes={scale=0.75}, at = {(1.2,-0.6)}, legend columns=4, anchor=south},
width=0.26\textwidth, grid=both, name=ax1]
\pgfplotstableread{plot/sated_noise.txt}\mydata;
\addplot[thick, mark=o] table
[x expr=\thisrow{w},y expr=\thisrow{pre}] {\mydata};
\addlegendentry{Precision};
\addplot[thick, mark=triangle] table
[x expr=\thisrow{w},y expr=\thisrow{rec}] {\mydata};
\addlegendentry{Recall};
\addplot[thick, mark=square] table
[x expr=\thisrow{w},y expr=\thisrow{auc}] {\mydata};
\addlegendentry{AUC};
\addplot[thick, dashed,mark=*] table
[x expr=\thisrow{w},y expr=\thisrow{acc}] {\mydata};
\addlegendentry{Accuracy};
\end{axis}

\begin{axis}[title style={yshift=-1.5ex}, title=\textbf{\Dialogue}, xlabel=Fraction of noise data, ymin=0., xmin=0.1, xmax=0.5, xtick=data,
width=0.26\textwidth, grid=both, at={(ax1.south east)}, xshift=1cm,]
\pgfplotstableread{plot/cornell_noise.txt}\mydata;
\addplot[thick, mark=o] table
[x expr=\thisrow{w},y expr=\thisrow{pre}] {\mydata};
\addplot[thick, mark=triangle] table
[x expr=\thisrow{w},y expr=\thisrow{rec}] {\mydata};
\addplot[thick, mark=square] table
[x expr=\thisrow{w},y expr=\thisrow{auc}] {\mydata};
\addplot[thick, dashed,mark=*] table
[x expr=\thisrow{w},y expr=\thisrow{acc}] {\mydata};
\end{axis}
\end{tikzpicture} 
\caption{Effect of noise and errors.}  
\label{fig:noise_data}
\end{figure}

\paragraphbe{Effect of noise and errors in the queries.}
$\Duser$ may be noisy or partially erroneous (e.g., if not all of $\Duser$
was used to train the target model $f$).  To evaluate how this affects
auditing, for each training user, we use part of his data to train $f$
and hold out the remaining fraction to represent noise during auditing.
We vary this fraction between 0.1, 0.2, \ldots, 0.5.

Fig.~\ref{fig:noise_data} shows the results. For SATED and \Dialogue,
recall drops significantly, close to 0 for SATED when the fraction
of noise is 0.5.  Increasing the amount of noise biases the audit
model towards misclassifying most training users as ``non-members.''
Precision and AUC remain high when noise increases.  This may indicate
that the scores of the membership classifier at the heart of the audit
model still have a distinguishable gap between members and non-members,
which is however not learned from the outputs of the shadow models
queried with clean data (see Section~\ref{sec:shadow}).


\begin{table}[t]
\caption{Examples of texts obfuscated using Google translation API and
Yandex translation API.}
\centering
\footnotesize
\begin{tabularx}{0.48\textwidth}{|X|} 
\hline
\textbf{No obfuscation:} i see so many adults that could benefit from this going around having themselves a big fat sugar snack or soda pop as a treat it 's so sad 
\\ \rule{0.45\textwidth}{0.4pt}
\textbf{Google:} i saw so many adults who can benefit from cherishing big fat sugar snacks and soda pop and going around, it is very sad
\\ \rule{0.45\textwidth}{0.4pt}
\textbf{Yandex:} i think a lot of adults have benefited over your big fat candy and and handling of grief
\\ \hline 
\end{tabularx}
\label{tbl:obfuscated}
\end{table}

\begin{table}[t]
\caption{Audit performance on obfuscated Reddit comments.}
\label{tbl:obfuscated_result}
\centering
\footnotesize
\begin{tabular}{l|rrrr}
     Dataset & Accuracy & AUC & Precision & Recall \\ \hline\hline
     Baseline & 1.000 & 1.000 & 1.000 & 1.000 \\
     Google & 0.580 & 0.858 & 0.944 & 0.170 \\
     Yandex & 0.500 & 0.782 & 0.500 & 0.010 \\ \hline 
\end{tabular}
\end{table}

\paragraphbe{Auditing obfuscated data.}
Finally, we evaluate the effect of obfuscation on the success of auditing.
This is the first step towards determining whether text-generation
models memorize specific word sequences (which would not be preserved
by obfuscation) rather than higher-level linguistic features (which
might be).

We use an obfuscation technique, previously considered for
evading author attribution~\cite{brennan2012adversarial}, that
machine-translates the text to a different language and back.
We obfuscate the training and test users' Reddit comments using
Google\footnote{\url{https://cloud.google.com/translate/}}
and Yandex\footnote{\url{https://tech.yandex.com/translate/}}
translation APIs to translate English to Japanese and back to English.
Table~\ref{tbl:obfuscated} shows examples of obfuscated text.

Table~\ref{tbl:obfuscated_result} reports the results of auditing
on obfuscated texts.  For both Google- and Yandex-based obfuscation,
audit accuracy drops to near random and recall is very low.  AUC scores
are still around 0.8, which is much higher than random guessing.
This indicates there is some useful signal in the model's outputs on
obfuscated texts, but the auditor's membership classifier\textemdash
which was trained on non-obfuscated texts\textemdash fails to capture
this signal.

This is a remarkable result given the poor quality of translation.
Even if the user's text has been garbled almost to the point of
incomprehensibility, in some cases there is still enough information
left to detect its presence in the training data.

\section{Memorization in text-generation models}
\label{sec:overfit}

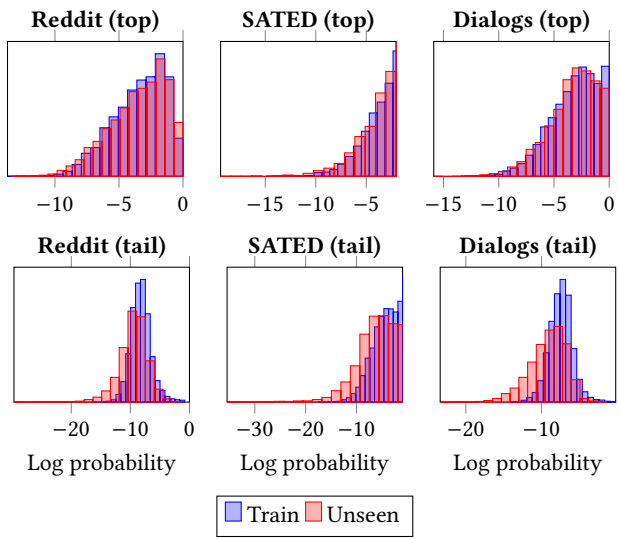
\begin{figure}[t!]
\centering
\begin{tabular}{c}
\begin{tikzpicture}
\begin{axis}[ybar, ytick=\empty, width=0.22\textwidth, enlarge x limits=false, ymin=0., title style={yshift=-1.5ex}, title=\textbf{Reddit (top)}, legend style={at={(0.01,0.8)}, anchor=west},
name=ax1]
\pgfplotstableread{plot/reddit_probs_0_1000.txt}\mydata;
\addplot[ybar interval, draw=blue, fill=blue, fill opacity=0.3] table [x expr=\thisrow{b1},y expr=\thisrow{n1}] \mydata;
\addplot[ybar interval, draw=red, fill=red, fill opacity=0.3] table [x expr=\thisrow{b2},y expr=\thisrow{n2}] \mydata;
\end{axis}

\begin{axis}[ybar, ytick=\empty, width=0.22\textwidth, enlarge x limits=false, ymin=0., ymax=0.1, title style={yshift=-1.5ex}, title=\textbf{SATED (top)}, legend style={at={(0.01,0.8)}, anchor=west},
name=ax2, at={(ax1.south east)}, xshift=0.5cm]
\pgfplotstableread{plot/sated_probs_0_1000.txt}\mydata;
\addplot[ybar interval, draw=blue, fill=blue, fill opacity=0.3] table [x expr=\thisrow{b1},y expr=\thisrow{n1}] \mydata;
\addplot[ybar interval, draw=red, fill=red, fill opacity=0.3] table [x expr=\thisrow{b2},y expr=\thisrow{n2}] \mydata;
\end{axis}

\begin{axis}[ybar, ytick=\empty, width=0.22\textwidth, enlarge x limits=false, ymin=0., ymax=0.2, xmax=0., title style={yshift=-1.5ex}, title=\textbf{Dialogs (top)}, legend style={at={(0.01,0.8)}, anchor=west}, at={(ax2.south east)}, xshift=0.5cm]
\pgfplotstableread{plot/cornell_probs_0_1000.txt}\mydata;

\addplot[ybar interval, draw=blue, fill=blue, fill opacity=0.3] table [x expr=\thisrow{b1},y expr=\thisrow{n1}] \mydata;
\addplot[ybar interval, draw=red, fill=red, fill opacity=0.3] table [x expr=\thisrow{b2},y expr=\thisrow{n2}] \mydata;
\end{axis}
\end{tikzpicture} 
\\
\begin{tikzpicture}
\begin{axis}[ybar, xlabel={Log probability}, ytick=\empty, width=0.22\textwidth, enlarge x limits=false, ymin=0., xmax=0., title style={yshift=-1.5ex}, title=\textbf{Reddit (tail)}, legend style={at={(0.01,0.8)}, anchor=west}, name=ax1]
\pgfplotstableread{plot/reddit_probs_1000_5000.txt}\mydata;
\addplot[ybar interval, draw=blue, fill=blue, fill opacity=0.3] table [x expr=\thisrow{b1},y expr=\thisrow{n1}] \mydata;
\addplot[ybar interval, draw=red, fill=red, fill opacity=0.3] table [x expr=\thisrow{b2},y expr=\thisrow{n2}] \mydata;
\end{axis}

\begin{axis}[ybar, xlabel={Log probability}, ytick=\empty, width=0.22\textwidth, enlarge x limits=false, ymin=0., ymax=0.1, title style={yshift=-1.5ex}, title=\textbf{SATED (tail)}, 
legend style={at={(0.5,-1.)}, anchor=south, legend columns=2},
name=ax2, at={(ax1.south east)}, xshift=0.5cm]
\pgfplotstableread{plot/sated_probs_1000_5000.txt}\mydata;

\addplot[ybar interval, draw=blue, fill=blue, fill opacity=0.3] table [x expr=\thisrow{b1},y expr=\thisrow{n1}] \mydata;
\addplot[ybar interval, draw=red, fill=red, fill opacity=0.3] table [x expr=\thisrow{b2},y expr=\thisrow{n2}] \mydata;
\legend{Train, Unseen}
\end{axis}

\begin{axis}[ybar, xlabel={Log probability}, ytick=\empty, width=0.22\textwidth, enlarge x limits=false, ymin=0., title style={yshift=-1.5ex}, title=\textbf{Dialogs (tail)}, legend style={at={(-0.5,-0.8)}, anchor=west, legend columns=2},
at={(ax2.south east)}, xshift=0.5cm]
\pgfplotstableread{plot/cornell_probs_1000_5000.txt}\mydata;
\addplot[ybar interval, draw=blue, fill=blue, fill opacity=0.3] table [x expr=\thisrow{b1},y expr=\thisrow{n1}] \mydata;
\addplot[ybar interval, draw=red, fill=red, fill opacity=0.3] table [x expr=\thisrow{b2},y expr=\thisrow{n2}] \mydata;
\end{axis}
\end{tikzpicture}
\end{tabular}
\caption{Histograms of log probabilities of words generated by our
text-generation models.  The top row are the histograms for the
top 20\% most frequent words, the bottom row are the histograms for
the rest.}
\label{fig:logprob}
\end{figure}


In this section, we analyze why auditing works so well for text-generation
models that are not overfitted as measured by their test-train accuracy
gap (see Section~\ref{sec:target_performance}).

\paragraphbe{Word frequency and probability.}
The loss function for the text-generation models is the sum of the
negative log probabilities of the words in the input sequence (see
Section~\ref{sec:back-language}).  By its very construction, this loss
function ``encourages'' the model to memorize sequences that occur in
the training data.


Fig.~\ref{fig:logprob} shows the histograms of the log probabilities
of the more and less frequent words in the training (``train'')
and test (``unseen'') sequences.  For the more frequent words, the
histograms for the training and test sequences are almost identical.
For the less frequent words, the model fits worse for both the training
and test sequences as modes focus on smaller log probability values.
Most importantly, there is a gap between the less frequent words in the
training sequences and those in the test sequences.  This gap indicates
that the model assigns higher probabilities to words in the training
sequences, producing a strong signal that can be used for membership
inference and consequently auditing.

These histograms also demonstrate that our text-generation models are
not overfitted to their training datasets in terms of the loss value.
The 20\% most frequent words account for 86.9\% of the training data and
88.1\% of the test data in Reddit, 89.5\% and 90.4\% in SATED, and 93.1\%
and 94.1\% in \Dialogue.  Consequently, these words dominate the training
and test loss value.  Not surprisingly, text-generation models typically
generate words from the top 20\% of the word-frequency distribution.
As long as the log probabilities remain similar for the top 20\% words
in both the training and test datasets, the training and test losses of
the model will be similar.

\begin{figure*}[htbp!]
\centering
\begin{tabular}{ccc}
\begin{tikzpicture}
\begin{axis}[title style={yshift=-1.5ex}, title=\textbf{Reddit}, xlabel=Frequency rank,ylabel= Predicted rank, xmin=0, xmax=4900, ymin=0, ymax=4900, width=0.35\textwidth,
legend style = { at = {(0.02,0.8)}, legend columns=1, anchor=west},
grid=both, yticklabel style={rotate=90}]
\pgfplotstableread{plot/reddit_rank_shift.txt}\mydata;
\addplot[thick,dashed,line legend] coordinates {(0, 0) (4900, 4900)};
\addplot[thick, mark=*, blue, mark size=1pt,line legend] table
[x expr=\thisrow{x},y expr=\thisrow{s1}] {\mydata};
\addplot+[name path=A, smooth, blue!30, no markers] table
[x expr=\thisrow{x},y expr=\thisrow{s11},forget plot] {\mydata};
\addplot+[name path=B, smooth, blue!30, no markers] table
[x expr=\thisrow{x},y expr=\thisrow{s12},forget plot] {\mydata};
\addplot[blue!30,forget plot] fill between[of=A and B];
\addplot[thick, mark=triangle*, red, mark size=1pt, line legend] table
[x expr=\thisrow{x},y expr=\thisrow{s2}] {\mydata};
\addplot+[name path=A, smooth, red!30, no markers,forget plot] table
[x expr=\thisrow{x},y expr=\thisrow{s21}] {\mydata};
\addplot+[name path=B, smooth, red!30, no markers,forget plot] table
[x expr=\thisrow{x},y expr=\thisrow{s22}] {\mydata};
\addplot[red!30,forget plot] fill between[of=A and B];
\legend{Frequency, Train, Unseen};
\end{axis}
\end{tikzpicture} 
&
\begin{tikzpicture}
\begin{axis}[title style={yshift=-1.5ex}, title=\textbf{SATED}, xlabel=Frequency rank, xmin=0, xmax=4900, ymin=0, ymax=4900, width=0.35\textwidth,
legend style = { at = {(0.02,0.8)}, legend columns=1, anchor=west}, grid=both, yticklabel style={rotate=90}]
\pgfplotstableread{plot/sated_rank_shift.txt}\mydata;
\addplot[thick,dashed,line legend] coordinates {(0, 0) (4900, 4900)};
\addplot[thick, mark=*, blue, mark size=1pt,line legend] table
[x expr=\thisrow{x},y expr=\thisrow{s1}] {\mydata};
\addplot+[name path=A, smooth, blue!30, no markers] table
[x expr=\thisrow{x},y expr=\thisrow{s11},forget plot] {\mydata};
\addplot+[name path=B, smooth, blue!30, no markers] table
[x expr=\thisrow{x},y expr=\thisrow{s12},forget plot] {\mydata};
\addplot[blue!30,forget plot] fill between[of=A and B];
\addplot[thick, mark=triangle*, red, mark size=1pt, line legend] table
[x expr=\thisrow{x},y expr=\thisrow{s2}] {\mydata};
\addplot+[name path=A, smooth, red!30, no markers,forget plot] table
[x expr=\thisrow{x},y expr=\thisrow{s21}] {\mydata};
\addplot+[name path=B, smooth, red!30, no markers,forget plot] table
[x expr=\thisrow{x},y expr=\thisrow{s22}] {\mydata};
\addplot[red!30,forget plot] fill between[of=A and B];
\legend{Frequency, Train, Unseen};
\end{axis}
\end{tikzpicture} 
&
\begin{tikzpicture}
\begin{axis}[title style={yshift=-1.5ex}, title=\textbf{\Dialogue}, xlabel=Frequency rank, xmin=0, xmax=4900, ymin=0, ymax=4900, width=0.35\textwidth,
legend style = { at = {(0.02,0.8)}, legend columns=1, anchor=west}, grid=both, yticklabel style={rotate=90}]
\pgfplotstableread{plot/cornell_rank_shift.txt}\mydata;
\addplot[thick,dashed,line legend] coordinates {(0, 0) (4900, 4900)};
\addplot[thick, mark=*, blue, mark size=1pt,line legend] table
[x expr=\thisrow{x},y expr=\thisrow{s1}] {\mydata};
\addplot+[name path=A, smooth, blue!30, no markers] table
[x expr=\thisrow{x},y expr=\thisrow{s11},forget plot] {\mydata};
\addplot+[name path=B, smooth, blue!30, no markers] table
[x expr=\thisrow{x},y expr=\thisrow{s12},forget plot] {\mydata};
\addplot[blue!30,forget plot] fill between[of=A and B];
\addplot[thick, mark=triangle*, red, mark size=1pt, line legend] table
[x expr=\thisrow{x},y expr=\thisrow{s2}] {\mydata};
\addplot+[name path=A, smooth, red!30, no markers,forget plot] table
[x expr=\thisrow{x},y expr=\thisrow{s21}] {\mydata};
\addplot+[name path=B, smooth, red!30, no markers,forget plot] table
[x expr=\thisrow{x},y expr=\thisrow{s22}] {\mydata};
\addplot[red!30,forget plot] fill between[of=A and B];
\legend{Frequency, Train, Unseen};
\end{axis}
\end{tikzpicture} 
\end{tabular}
\caption{Ranks of words in the frequency table of the training corpus
and in the models' predictions (lower rank means that the word is
more likely).  
Shaded area is the 95\% confidence interval for all occurrences of the
word in the data.  These charts demonstrate that the models assign much
higher rank to words when they appear in training sequences vs.\ when
they appear in test sequences, especially for the less-frequent words.} 
\label{fig:rankshift} 
\end{figure*}
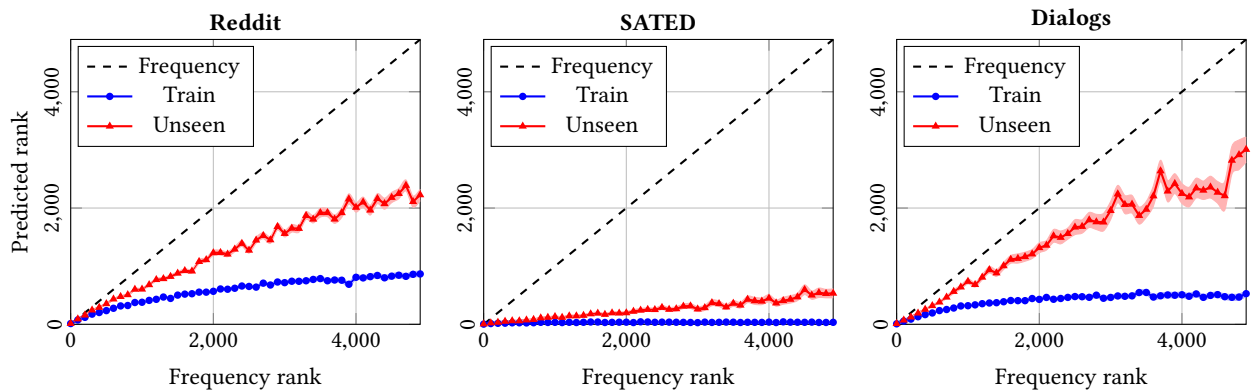

\paragraphbe{Word frequency and predicted rank.}
Memorization of training sequences produces a much stronger signal in
the relative \emph{rank} assigned by the model to the candidate words
in the model's output vocabulary.  Fig.~\ref{fig:rankshift} shows
the relationship between a word's rank in the frequency table of the
training corpus and its rank in the model's predictions.  A smaller rank
number indicates that the word is ranked higher in the vocabulary, i.e.,
more frequent in the corpus or more likely to be predicted by the model.
On all datasets, less frequent words exhibit a much bigger gap between
the rank predicted by the model when the word appears in a training
sequence and when it appears in a test sequence.  This explains why
our auditing algorithm is more successful when it queries the target
model with sequences consisting of the less-frequent words (see
Section~\ref{sec:selection}).

\begin{figure}[t!]
\centering
\begin{tikzpicture}
\begin{axis}[title style={yshift=-1.5ex}, title=\textbf{Reddit}, xlabel=\% hidden units ablated, ylabel=Training accuracy, 
xmin=0., xmax=0.5, xtick={0., 0.1, 0.2, 0.3, 0.4, 0.5}, 
legend style = {nodes={scale=0.75}, at = {(1.1,-0.6)}, legend columns=2, anchor=south},
width=0.26\textwidth, grid=both, name=ax1]
\pgfplotstableread{plot/reddit_ablation.txt}\mydata;
\addplot[thick, mark=o] table
[x expr=\thisrow{p},y expr=\thisrow{b1}] {\mydata};
\addlegendentry{Top 10\% words};
\addplot[thick, mark=triangle] table
[x expr=\thisrow{p},y expr=\thisrow{b2}] {\mydata};
\addlegendentry{Tail 90\% words};
\end{axis}
\begin{axis}[title style={yshift=-1.5ex}, title=\textbf{SATED}, xlabel=\% hidden units ablated, xmin=0., xmax=0.9,  xtick={0., 0.1, 0.3, 0.5, 0.7, 0.9},
width=0.26\textwidth, grid=both, at={(ax1.south east)}, xshift=1cm]
\pgfplotstableread{plot/sated_ablation.txt}\mydata;
\addplot[thick, mark=o] table
[x expr=\thisrow{p},y expr=\thisrow{b1}] {\mydata};
\addplot[thick, mark=triangle] table
[x expr=\thisrow{p},y expr=\thisrow{b2}] {\mydata};
\end{axis}
\end{tikzpicture} 
\caption{Ablation analysis on Reddit and SATED.}
\label{fig:ablation}
\end{figure}
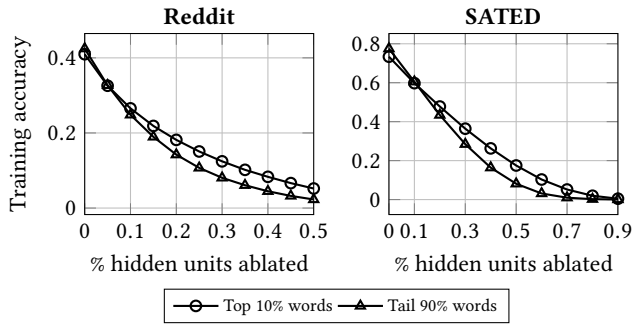

\paragraphbe{Ablation analysis.}
We have shown that probabilities and ranks produced by text-generation
models exhibit a gap between the training and test sequences for the
less-frequent words but not for the most-frequent words.  We hypothesize
that these models learn generalizable patterns for the most-frequent words
while hard-memorizing the sequences consisting of the less-frequent words.

To gather evidence for this hypothesis, we carried out an experiment based
on ablation analysis that was recently proposed to detect memorization
in deep-learning models~\cite{morcos2018importance}.  As more hidden
units are ablated, accuracy on the training data degrades quicker for
models that are hard-memorizing the training data.


We train target models without dropout (since dropout ablates the
hidden units during training) on Reddit and SATED, keeping the other
hyper-parameters the same as in Section~\ref{sec:setup}.  We randomly set
a fraction of the model's hidden representations to zero and evaluate the
accuracy of word prediction on the training data.  We vary the fraction
from 0.1 to 0.5 on Reddit and 0.1 to 0.9 on SATED and report the accuracy
score separately for the 10\% most frequent words and the remaining 90\%
in Fig.~\ref{fig:ablation}.

When no hidden units are ablated, accuracy is similar for the
most-frequent words and the rest.  As the fraction of ablated units
increases, accuracy on the less-frequent words drops more significantly
than on the most-frequent words.  This indicates that predicting
less-frequent words is more dependent on specific hidden units in the
model and thus involves more memorization.

\section{Limitations of auditing}
\label{sec:limit}

\paragraphbe{Models trained on a very large number of users.}
In some industrial implementation of text-generation
models~\cite{mcmahan2017communication, mcmahan2017learning},
the number of users is on the scale of millions.  Performance of
our auditor starts to drop when the number of users reaches 10,000
(Section~\ref{sec:selection}).  We expect that our black-box algorithm
will not be able to audit models trained on a very large number (dozens
or hundreds of thousands) of users.  That said, (a) many state-of-the-art
models are trained on fewer than 10,000 users~\cite{michel2018extreme,
kottur2017exploring, zhang2018personalizing}, and (b) white-box auditing
techniques may be effective even against models trained on dozens of
thousands of users.  This is a topic for future work.


\paragraphbe{Deeper models.} 
In our experiments, both the target and shadow models are one-layer
LSTMs or GRUs.  We have not experimented with auditing deeper and
more sophisticated models.  We expect that such models are even more
susceptible to memorization, but this is another topic for future
research.

\paragraphbe{Differentially private models.} 
In theory, user-level differential privacy (DP) is a direct countermeasure
to user-level membership inference.  We used federated learning with
differential privacy~\cite{mcmahan2017learning} to train a next-word
prediction model on the Reddit dataset, setting the number of users
to 5,000, user sampling rate to 0.04 per round, $L_2$ bound on a
single user's contribution to 10.0, and the other hyper-parameters
as in~\cite{mcmahan2017learning}.  After 300 rounds of training,
this produced an ($\epsilon, \delta$)-DP model with $\epsilon=4.129$
and $\delta=1e-4$ which achieves 15\% word prediction accuracy,
similar to~\cite{mcmahan2017learning}.  By contrast, the accuracy of
our non-DP model is 20\% when trained on only 100 users, i.e., the DP
model is significantly less accurate than the non-DP one.  Our auditing
algorithm fails against the DP model, with performance scores near 0.5
(equivalent to random guessing).

\begin{figure}[t!]
\centering
\begin{tikzpicture}
\begin{axis}[title style={yshift=-1.5ex}, title=\textbf{DP word prediction on Reddit}, xlabel=Frequency rank,ylabel= Predicted rank, xmin=0, xmax=4900, ymin=0, ymax=4900, width=0.35\textwidth,
legend style = {at = {(0.02,0.8)}, legend columns=1, anchor=west}, grid=both]
\pgfplotstableread{plot/reddit_dp_rank_shift.txt}\mydata;
\addplot[thick,dashed,line legend] coordinates {(0, 0) (4900, 4900)};
\addplot[thick, mark=*, blue, mark size=1pt,line legend] table
[x expr=\thisrow{x},y expr=\thisrow{s1}] {\mydata};
\addplot+[name path=A, smooth, blue!30, no markers] table
[x expr=\thisrow{x},y expr=\thisrow{s11},forget plot] {\mydata};
\addplot+[name path=B, smooth, blue!30, no markers] table
[x expr=\thisrow{x},y expr=\thisrow{s12},forget plot] {\mydata};
\addplot[blue!30,forget plot] fill between[of=A and B];
\addplot[thick, mark=triangle*, red, mark size=1pt, line legend] table
[x expr=\thisrow{x},y expr=\thisrow{s2}] {\mydata};
\addplot+[name path=A, smooth, red!30, no markers,forget plot] table
[x expr=\thisrow{x},y expr=\thisrow{s21}] {\mydata};
\addplot+[name path=B, smooth, red!30, no markers,forget plot] table
[x expr=\thisrow{x},y expr=\thisrow{s22}] {\mydata};
\addplot[red!30,forget plot] fill between[of=A and B];
\legend{Frequency, Train, Unseen};
\end{axis}
\end{tikzpicture} 
\caption{Ranks of words in the training corpus and in the predictions
of the differentially private model.}
\label{fig:dp_rankshift}
\end{figure}
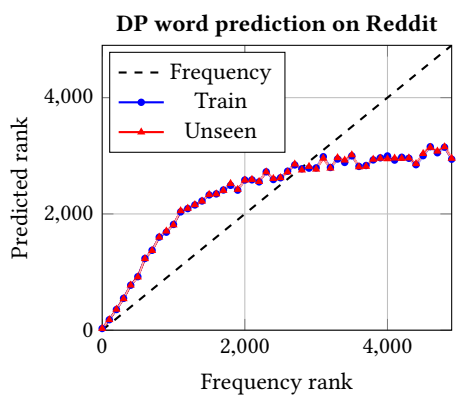

To further investigate the predictive power of the DP model,
Fig.~\ref{fig:dp_rankshift} plots the ranks of words in the vocabulary
(based on their frequencies) and in the model's predictions.
The predicted rank is larger than the frequency rank for the 50\% most
frequent words and remains around 3,000 for the other 50\%.  The predicted
rank is very similar for the words in the training and test sequences,
which explains why auditing fails.

The plot also suggests that the differentially private model will almost
always predict common words and hardly ever predict relatively rare words.
While it does not appear that the model memorizes its training data,
it is not clear to what extent it generalizes.

\section{Related work}
\label{sec:related}

\paragraphbe{Membership inference.}  
Membership inference attacks involve observing the output of some
computations over a hidden dataset $\D$ and determining whether
a specific data point is a member of $\D$.  Membership inference
attacks against aggregate statistics have been demonstrated in the
context of genomic studies~\cite{homer2008resolving}, location
time-series~\cite{pyrgelis2017knock}, and noisy statistics in
general~\cite{dwork2015robust}.


Shokri et al.~\cite{shokri2017membership} develop black-box
membership inference techniques against ML models which perform
best when the target model is overfitted to the training data.
Truex et al.~\cite{truex2018towards} extend and generalize this
work to white-box and federated-learning settings.  Rahman et
al.~\cite{rahman2018membership} use membership inference to
evaluate the tradeoff between test accuracy and membership privacy in
differentially private ML models.  Hayes et al.~\cite{hayes2017logan}
study membership inference against generative models.  Long et
al.~\cite{long2018understanding} show that well-generalized models
can leak membership information, but the adversary must first
identify a handful of vulnerable records in the training dataset.
Yeom et al.~\cite{yeom2018privacy} formalize membership inference and
theoretically show that overfitting is sufficient but not necessary.

\paragraphbe{Memorization in ML models.} 
Zhang et al.~\cite{zhang2016understanding} show that deep learning models
can achieve perfect accuracy even on randomly labeled training data.
Song et al.~\cite{song2017machine} present algorithms that intentionally
encode the training data in the model.  By contrast, we demonstrate that
popular text-generation models \emph{unintentionally} memorize their
training data.


Carlini et al.~\cite{carlini2018secret} show that a black-box adversary
can extract specific \emph{numbers} that occur in the training data
of a generative model, given some prior knowledge about the format
(e.g., a credit card number).  For a text-generation model, numbers are
essentially random data, thus this is another illustration that models
memorize random data.  By contrast, we show that text-generation models
memorize even words and sentences that are directly related to their
primary task and leverage this into an effective auditing method.



\paragraphbe{User-level differential privacy.}
User-level differential privacy (DP) bounds the influence of any single
user on the model.  McMahan et al.\ propose a DP federated learning
algorithm for language models~\cite{mcmahan2017learning}.  With the
current state of the art, a massive number of users (at least 10,000)
is needed to create DP models that achieve reasonable accuracy.  How to
build accurate DP models with fewer users remains an open question.


\paragraphbe{Auditing ML models.} 
Much recent work aims to understand the behavior of ML models with
black-box access~\cite{koh2017understanding, adler2018auditing}.
These approaches improve the interpretability of the model by
showing how features or training data points influence the model's
predictions.  Other model-auditing research focuses on detecting bias
and discrimination~\cite{tan2017detecting, tramer2017fairtest}.  We are
not aware of any prior work that aims to audit the use of specific data
sources to train a model.


\section{Conclusion}

Deep learning-based, text-generation models for word prediction,
translation, and dialog generation are core components of many popular
online services.  We demonstrated that these models memorize their
training data.  This memorization does not appear to manifest in reduced
test accuracy, which is a symptom of ``conventional'' overfitting, but
is reflected instead in how they rank the candidate words they generate.


We developed a black-box auditing method that enables users to check
if their chats, messages, or comments have been used to train someone
else's model.  Our auditing method, based on a new flavor of membership
inference that exploits memorization in text-generation models, is
very effective.  More powerful auditing algorithms may be possible if
the auditor has access to the model's parameters and can observe its
internal representations rather than just output predictions.  This is
a topic for future work.


We view the results of this paper as essentially positive, demonstrating
how memorization in ML models can help detect unauthorized uses of
sensitive personal data and ensure compliance with GDPR and other
data-protection policies and regulations.

\paragraphbe{Acknowledgments.}
Supported in part by the NSF grants 1611770 and 1704296 and the generosity
of Eric and Wendy Schmidt by recommendation of the Schmidt Futures program

\bibliographystyle{abbrv}
\bibliography{citation}

\appendix
\twocolumn[
\begin{@twocolumnfalse}
\vbox{\parbox[t]{\textwidth}{
\centering
\Huge\sffamily\bfseries Reproducibility Information
\vspace{0.5em}}}
\end{@twocolumnfalse}
]

In this appendix, we provide the pseudo-code of the auditing algorithm
and describe all model architectures, configurations, and hyper-parameters
needed to reproduce the results.

\section{Pseudo-code for auditing}

Algorithm~\ref{alg:overview} is the pseudo-code for the auditing
process.  Function~\texttt{AuditMembership} is used to audit if
the user's data $\Duser$ was included in the training data of the
target model $f$.  The auditor first trains an audit model using
function~\texttt{TrainAuditModel}.  He then extracts histogram features
$h_u$ on (possibly sampled) user's data using function \FeatExtract.
Finally, the audit model predicts if the user's data was used to train
$f$ given the extracted features $h_u$.

Function \texttt{TrainAuditModel} is the procedure for training the
audit model.  The auditor first trains $k$ shadow models with a randomly
sampled set of users, collects their outputs, extracts feature vectors for
the training and test users, and labels these feature vectors accordingly.
The auditor then trains a binary classifier based on the labeled feature
vectors.

Function \FeatExtract extracts features from the predicted ranks, as
described in Section~\ref{sec:audittrain}.  The feature vector is a
$d$-dimensional vector representing the histogram of predicted ranks.
The $i$th entry of the vector is the count of ranks in the range
$[(i-1)\cdot b, i\cdot b]$, where $b$ is the histogram bin size.

When the number of queries to the target model is limited to $m$, the
auditor uses function \texttt{SampleQueries} to sample a subset of the
user's data.  The auditor can choose either $m$ texts at random, or $m$
texts with the smallest word frequencies.

\begin{algorithm}[!t]
\caption{\normalsize Auditing text-generation models}
\begin{algorithmic}
\State \textbf{Hyper-parameters:} auditor's reference dataset $\Dref$,
number of shadow models $k$, user's data $\Duser$, target model
$f$, target model-training protocol $\mathcal{T}_\text{target}$, audit
model-training protocol $\mathcal{T}_\text{audit}$, maximum number of
queries $m$, number of bins in histogram $d$ 
\State 
\State \textbf{function} \texttt{AuditMembership}()
\Indent 
\State  $f_\text{audit}\gets$\texttt{TrainAuditModel}()
\State $\D_{\text{sample}, u}\gets$\Sample($m, \Duser$)
\State $h_u\gets$\FeatExtract$(f, \D_{\text{sample}, u})$
\State \Return prediction of membership $f_\text{audit}(h_u)$
\EndIndent 
\State 
\State \textbf{function} \Sample($m, \D$)
\Indent 
\If{random sample}
\State \Return randomly selected $m$ rows in $\D$
\Else \Comment{sample based on frequency}
\State $C\gets$\{$\Sigma$ (frequency of $w$ for $w$ in $y$) $| \forall (x, y)\in\D$\}
\State $I\gets$ indices of $m$ smallest values in $C$ 
\State \Return $m$ rows in $\D$ indexed by $I$
\EndIf
\EndIndent 
\State 
\State \textbf{function} \texttt{TrainAuditModel}()
\Indent 
\State $\D_\text{audit}\gets\emptyset$ \Comment{dataset for building the audit model}
\State $\Uref\gets$ users in $\Dref$
\For{$i=1$ to $k$} \Comment{train $k$ shadow models}
\State $\U_\text{ref}^\text{train}, \U_\text{ref}^\text{test}\gets$ random split $\Uref$
\State $\Dref^{\text{train}}\gets \cup_{u\in\U_\text{ref}^\text{train}}\{\D_{\text{ref}, u}\}$
\State Train a shadow model $f^\prime_i\gets\mathcal{T}_\text{target}(\Dref^\text{train})$.
\For{every $u$ in users of $\Uref$}
\State $\D_{\text{ref}, u}\gets$ data in $\Dref$ associated with $u$
\State $h^\prime_u\gets$\FeatExtract$(f^\prime_i, \D_{\text{ref}, u})$
\State $z^\prime_u\gets 1$ if $u$ in $\U_\text{ref-train}$ else 0 
\State $\D_\text{audit}\cup\{(h^\prime_u, z^\prime_u)\}$
\EndFor
\EndFor
\State Train the audit model $f_\text{audit}\gets\mathcal{T}_\text{audit}(\D_\text{audit})$
\State \Return $f_\text{audit}$
\EndIndent 
\State
\State \textbf{function} \FeatExtract($f$, $\D$)
\Indent
\State $R\gets\{\textbf{rank}(y)\text{ in }f(x)| \forall (x, y)\in \D\}$
\State Initialize feature vector $h$ with $d$ entries.
\State $b\gets |V| / d$ \Comment{histogram bin size} 
\For{$i=1$ to $d$} \Comment{count of ranks in each bin}
\State $h_i=|\{(i-1)\cdot b \leq r < i\cdot b | r\in R \}|$ %
\EndFor
\State \Return feature vector $h$
\EndIndent 
\end{algorithmic}
\label{alg:overview}
\end{algorithm}

\section{Experiment setup}

\subsection{Target models}
\label{sec:setup}

\paragraphbe{Next-word prediction.} 
We use a one-layer long short-term memory (LSTM)~\cite{hochreiter1997long}
as the target model.  LSTM is a more complicated RNN that can capture the
long-term dependency in the sequence.  The input sequence of tokens is
first mapped to a sequence of embeddings.  The embedding is then fed
to the LSTM that learns a hidden representation for the context for
predicting the next word.

\paragraphbe{Neural machine translation.} 
We use a sequence-to-sequence target model with the attention module as
described in~\cite{michel2018extreme}.  Both the encoder and the decoder
are one-layer LSTMs that operate on the embedding of source tokens
and target tokens.  The attention module adds an additional layer that
operates on all hidden representations in the encoder LSTM and helps
the decoder determine where to pay attention in the source texts when
predicting a token in the target language.

\paragraphbe{Dialog generation.} 
We use a sequence-to-sequence model without the attention module.
The encoder and the decoder are one-layer LSTMs.

\subsection{Hyper-parameters}
\label{sec:config}

\paragraphbe{Target models.}
We train the word-prediction model on the comments of 300 randomly
selected users from the Reddit dataset.  We set both the embedding
dimension and LSTM hidden-representation size to 128.  For training the
LSTM, we use the Adam optimizer~\cite{kingma2014adam} with the learning
rate set to 1e-3, batch size to 35, and the number of training epochs
to 30.

We train the translation and dialog-generation models on 300 randomly
selected users from SATED and \Dialogue, respectively.  We set both the
embedding dimension and LSTM hidden-representation size in the encoder
and decoder to 128.  We use the Adam optimizer with the learning rate
set to 1e-3, batch size to 20, and the number of training epochs to 30.

For all datasets, we fix the vocabulary to the most frequent 5,000 tokens
in the training texts.  Tokens not in the vocabulary are replaced with
a special $<$UNK$>$ token.  To prevent overfitting, we add dropout with
0.5 rate to all hidden layers of all models.


\paragraphbe{Shadow models.} 
For the experiments in Section~\ref{sec:diffhyper}, we construct shadow
models using different hyper-parameters than the target models.  On all
tasks, we used Gated Recurrent Units (GRU)~\cite{cho2014learning} instead
of LSTM.  The size of hidden units and embedding is set to 64, 96, 128,
160, \dots, 352 for the shadow models.  We optimize the shadow models
using momentum SGD with the learning rate set to 0.01, momentum set to
0.9, and number of training epochs to 50.

\subsection{Implementation}

\paragraphbe{Target and shadow models.} 
All target and shadow models were implemented
with Keras\footnote{\url{https://keras.io/}} using
TensorFlow~\cite{abadi2016tensorflow} backend.

\paragraphbe{Audit model.}  
We use linear SVM implemented in LIBLINEAR~\cite{fan2008liblinear}
to train the audit model with the default
hyper-parameters.\footnote{\url{http://scikit-learn.org/stable/modules/generated/sklearn.svm.LinearSVC.html}}

\paragraphbe{Hardware.} 
All models were trained on a machine with 3 NVIDIA Titan X GPUs, 8-core
Intel(R) Core(TM) i7-5960X CPU @ 3.00GHz and 94 GBs of RAM.

\end{document}